 \documentclass[final,5p,times,twocolumn,authoryear]{elsarticle}

\usepackage{amssymb}
\usepackage{lipsum}

\usepackage{graphicx} 
\usepackage{txfonts}
\usepackage{amsmath}	
\usepackage{xcolor,colortbl}
\usepackage{soul}
\usepackage{xspace}
\usepackage{enumerate}
\usepackage{placeins}

\usepackage[caption=false]{subfig}
\usepackage{pdflscape}

\usepackage{rotfloat}
\usepackage{adjustbox}

\usepackage{booktabs}

\usepackage{makecell}

\usepackage{multirow}
\usepackage{arydshln} 
\usepackage{natbib}
\bibpunct{(}{)}{;}{a}{}{,} 

\usepackage[usestackEOL]{stackengine}

\usepackage[pdftex,bookmarks=true,colorlinks=true,linkcolor=blue,citecolor=blue,breaklinks]{hyperref}      
\usepackage{scalerel}

\usepackage{tikz}

\usepackage{pgfplots}
\pgfplotsset{compat=1.17}

\newcommand{\ConfusionMatrixViridis}[4]{%
    \pgfmathsetmacro{\maxValue}{max(#1, #2, #3, #4)}

    \pgfmathsetmacro{\scaleTP}{round(9 * (#1 / \maxValue))}
    \pgfmathsetmacro{\scaleFP}{round(9 * (#2 / \maxValue))}
    \pgfmathsetmacro{\scaleFN}{round(9 * (#3 / \maxValue))}
    \pgfmathsetmacro{\scaleTN}{round(9 * (#4 / \maxValue))}

    \definecolor{viridis0}{rgb}{0.267,0.005,0.329}
    \definecolor{viridis1}{rgb}{0.283,0.141,0.458}
    \definecolor{viridis2}{rgb}{0.254,0.265,0.530}
    \definecolor{viridis3}{rgb}{0.207,0.372,0.553}
    \definecolor{viridis4}{rgb}{0.164,0.471,0.558}
    \definecolor{viridis5}{rgb}{0.128,0.566,0.551}
    \definecolor{viridis6}{rgb}{0.157,0.668,0.508}
    \definecolor{viridis7}{rgb}{0.369,0.788,0.382}
    \definecolor{viridis8}{rgb}{0.678,0.865,0.183}
    \definecolor{viridis9}{rgb}{0.993,0.906,0.144}

    \pgfmathtruncatemacro{\tpColor}{\scaleTP}
    \pgfmathtruncatemacro{\fpColor}{\scaleFP}
    \pgfmathtruncatemacro{\fnColor}{\scaleFN}
    \pgfmathtruncatemacro{\tnColor}{\scaleTN}

    \begin{tikzpicture}
        \fill[viridis\tpColor] (0,1) rectangle (1,2);
        \node at (0.5,1.5) {\textbf{#1}};

        \fill[viridis\fpColor] (1,1) rectangle (2,2);
        \node[text=white] at (1.5,1.5) {\textbf{#2}};

        \fill[viridis\fnColor] (0,0) rectangle (1,1);
        \node[text=white] at (0.5,0.5) {\textbf{#3}};

        \fill[viridis\tnColor] (1,0) rectangle (2,1);
        \node[text=white] at (1.5,0.5) {\textbf{#4}};

        \draw[thick] (0,0) rectangle (2,2);
        \draw[thick] (0,1) -- (2,1);
        \draw[thick] (1,0) -- (1,2);

        \node[above] at (0.5,2.) {\textbf{1}};
        \node[above] at (1.5,2.) {\textbf{0}};
        \node[left] at (-0.,1.5) {\textbf{1}};
        \node[left] at (-0.,0.5) {\textbf{0}};

        \node[rotate=90] at (-0.6,1) {\textbf{True}};

        \node[above] at (1,2.5) {\textbf{Predicted}};
    \end{tikzpicture}%
}

\usetikzlibrary{svg.path,fit}

\definecolor{orcidlogocol}{HTML}{A6CE39}
\tikzset{
  orcidlogo/.pic={
    \fill[orcidlogocol] svg{M256,128c0,70.7-57.3,128-128,128C57.3,256,0,198.7,0,128C0,57.3,57.3,0,128,0C198.7,0,256,57.3,256,128z};
    \fill[white] svg{M86.3,186.2H70.9V79.1h15.4v48.4V186.2z}
                 svg{M108.9,79.1h41.6c39.6,0,57,28.3,57,53.6c0,27.5-21.5,53.6-56.8,53.6h-41.8V79.1z M124.3,172.4h24.5c34.9,0,42.9-26.5,42.9-39.7c0-21.5-13.7-39.7-43.7-39.7h-23.7V172.4z}
                 svg{M88.7,56.8c0,5.5-4.5,10.1-10.1,10.1c-5.6,0-10.1-4.6-10.1-10.1c0-5.6,4.5-10.1,10.1-10.1C84.2,46.7,88.7,51.3,88.7,56.8z};
  }
}

\newcommand\orcidicon[1]{\href{https://orcid.org/#1}{\mbox{\scalerel*{
\begin{tikzpicture}[yscale=-1,transform shape]
\pic{orcidlogo};
\end{tikzpicture}
}{|}}}}
\bibpunct{(}{)}{;}{a}{}{,}             

\definecolor{scc}{rgb}{0.54, 0.17, 0.89}

\definecolor{gri}{rgb}{0.19, 0.55, 0.91}

\definecolor{yle}{rgb}{1.0, 0.16, 0.64}

\definecolor{sm}{rgb}{0.3, 0.4, 0.8}

\definecolor{cadmiumred}{rgb}{0.89, 0.0, 0.13}

\definecolor{darkpastelgreen}{rgb}{0.01, 0.75, 0.24}

\definecolor{lavender}{rgb}{0.9, 0.9, 0.98}

\definecolor{redi}{RGB}{255,38,0}
\definecolor{redii}{RGB}{200,50,30}
\definecolor{yellowi}{RGB}{255,251,0}
\definecolor{bluei}{RGB}{0,150,255}
\definecolor{blueii}{RGB}{135,247,210}
\definecolor{blueiii}{RGB}{91,205,250}
\definecolor{blueiv}{RGB}{115,244,253}
\definecolor{bluev}{RGB}{1,58,215}
\definecolor{orangei}{RGB}{240,143,50}
\definecolor{yellowii}{RGB}{222,247,100}
\definecolor{greeni}{RGB}{166,247,166}

\definecolor{greenii}{RGB}{18,177,191}
\definecolor{greeniii}{RGB}{170,213,220}
\definecolor{pinki}{RGB}{212,10,168}
\definecolor{violeti}{RGB}{118,42,131}

\newcommand{\um}{$\mu$m}

\newcommand{\lm}{\mbox{$L/M$}} 

\newcommand\arcdeg{\mbox{$^\circ$}}%
\newcommand\arcsec{\mbox{$''$}}%

\newcommand{\hg}{\mbox{Hi-GAL}}
\newcommand{\spitzer}{\mbox{\textit{Spitzer}}}
\newcommand{\gli}{\mbox{\textit{GLIMPSE}}}
\newcommand{\wise}{\mbox{\textit{WISE}}}

\newcommand{\uk}{\mbox{\textit{UKIDSS}}}
\newcommand{\twom}{\mbox{\textit{2MASS}}}

\newcommand{\STEvMSCARS}{\mbox{\texttt{YSO-Class}}\xspace}
\newcommand{\cp}{\mbox{\texttt{ClumpPopulator}}}
\newcommand{\evo}{\mbox{\texttt{EVO-STATS}}}

\defcitealias{maruccia}{M25}

\journal{Astronomy $\&$ Computing}

\begin{document}

\begin{frontmatter}



\title{The Evolutionary Path of Star-Forming Clumps in Hi-GAL}


\author[first]{Y.~Maruccia}
\author[first,second]{S.~Cavuoti}
\author[third,first]{M.~Brescia}
\author[first]{G.~Riccio}
\author[fourth]{S.~Molinari}
\author[fourth]{D.~Elia}
\author[fourth]{E.~Schisano}

\affiliation[first]{organization={INAF - Astronomical Observatory of Capodimonte},
            addressline={Via Moiariello 16}, 
            city={Napoli},
            postcode={I-80131}, 
            state={Italy},
            country={}}

\affiliation[second]{organization={INFN section of Naples},
            addressline={via Cinthia 6}, 
            city={Napoli},
            postcode={I-80126}, 
            state={Italy},
            country={}}
            
\affiliation[third]{organization={Department of Physics ``E. Pancini'', University Federico II of Napoli},
            addressline={Via Cinthia 21}, 
            city={Napoli},
            postcode={I-80126}, 
            state={Italy},
            country={}}
            
\affiliation[fourth]{organization={INAF - Istituto di Astrofisica e Planetologia Spaziale},
            addressline={Via Fosso del Cavaliere 100}, 
            city={Roma},
            postcode={I-00133}, 
            state={Italy},
            country={}}

\begin{abstract}
Star formation (SF) studies are benefiting from the huge amount of data made available by recent large-area Galactic plane surveys conducted between 2~$\mu$m and 3 mm. Fully characterizing SF demands integrating far-infrared/sub-millimetre (FIR/sub-mm) data, tracing the earliest phases, with near-/mid-infrared (NIR/MIR) observations, revealing later stages characterized by Young Stellar Objects (YSOs) just before main sequence star appearance. However, the resulting dataset is often a complex mix of heterogeneous and intricate features, limiting the effectiveness of traditional analysis in uncovering hidden patterns and relationships. In this framework, machine learning emerges as a powerful tool to handle the complexity of feature-rich datasets and investigate potential physical connections between the cold dust component traced by FIR/sub-mm emission and the presence of YSOs. We present a study on the evolutionary path of star forming clumps in the \hg\ survey through a multi-step approach, with the final aims of (a) obtaining a robust and accurate set of features able to well classify the star forming clumps in \hg\ based on their evolutionary properties, (b) establishing whether a connection exists between the cold material reservoir in clumps, traced by FIR/sub-mm emission, and the already formed YSOs, precursors of stars. For these purposes, our designed experiments aim at testing whether the FIR/sub-mm properties related to clumps are sufficient to predict the clump evolutionary stage, without considering the direct information about the embedded YSOs at NIR/MIR. Our machine learning-based method involves a four-step approach, based on feature engineering, data handling, feature selection and classification. 
This workflow ensures the identification of the most relevant features driving the SF process, and rigorously evaluates the results through a classification analysis. Our findings suggest that FIR/sub-mm and NIR/MIR emissions trace different evolutionary phases of star forming clumps, highlighting the complex and asynchronous nature of the SF process.
\end{abstract}




\end{frontmatter}




\section{Introduction}
\label{introduction}
\begin{figure*}
	\includegraphics[width=\hsize]{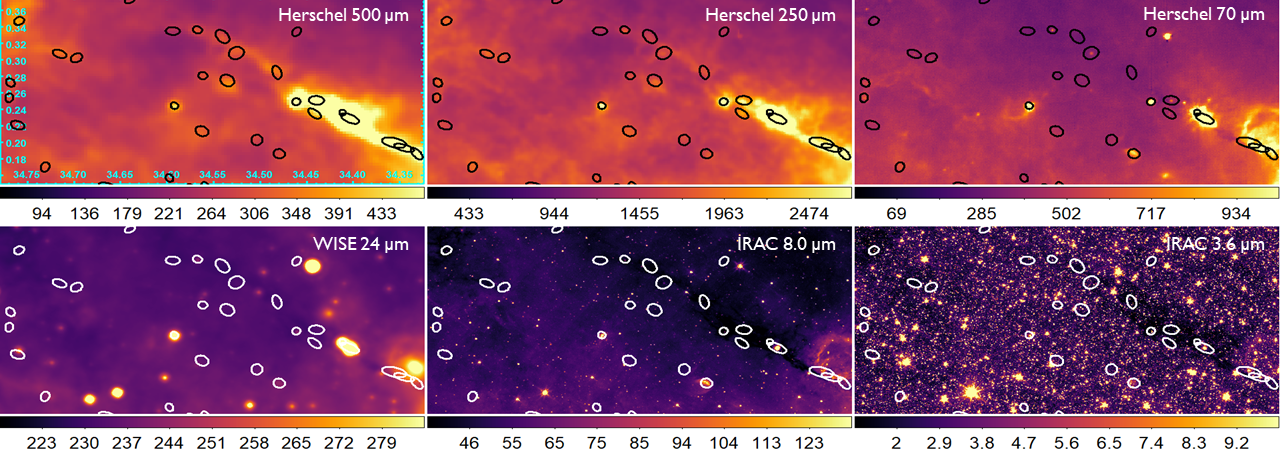}
    \caption{Portion of the same Galactic plane at different wavelengths. Ellipses are clumps from the \hg\ catalog at 350~$\mu$m, and they are superimposed on different images.}
    \label{fig:ds9}
\end{figure*}

Recent large-scale Galactic plane surveys conducted in the wavelength range from 2 $\mu$m and 3 mm have provided an invaluable wealth of data for the study of star formation processes. These surveys offer unprecedented insights into the birth and evolution of stars, thanks to the vast amount of information they have made available. By observing the Galactic plane across this broad spectrum, researchers are uncovering crucial details about the conditions, mechanisms, and environments that drive the formation of stars, enhancing our understanding of this phenomenon. Star formation occurs within vast and complex structures called giant molecular clouds (GMCs). These regions contain high concentrations of gas and dust and exhibit a network of interconnected clumps and filaments, where stars are born \citep{molinari2010clouds, ballesteros2020diffuse, chevance2022life}.
Observations have revealed that clumps within these clouds often serve as the breeding grounds for star clusters \citep[e.g.][]{elia21}. Within these clumps, smaller structures known as cores emerge through fragmentation processes \citep[e.g.][]{motte2022alma} and undergo gravitational collapse. Cores are where the formation of individual stars or bound systems of protostars takes place. Notably, young stellar objects (YSOs), which are still in their early stages of development, reside within these dusty cores \citep{mckee2003formation, beuther2002massive, wang2011hierarchical, wang2012protostellar, wang2014hierarchical, zhang2015fragmentation, pillai2019massive}. They are typically faint or even undetectable in the optical band, but their emissions, stemming from the contraction of proto-stellar cores and the surrounding dusty circumstellar material, become visible in the near-infrared (NIR), mid-infrared (MIR), and far-infrared (FIR) wavelengths \citep[e.g.][]{gutermuth2004ngc, whitney2004glimpse}. During the early Class 0 stage \citep{andre1993submillimeter}, the emission of a YSO is predominantly governed by a dense circumstellar envelope. This envelope efficiently reprocesses all the radiation from the deeply embedded forming protostar into the FIR regime. At this stage, the YSO is detectable only at wavelengths longer than 60--100~\um. As the system evolves, continued accretion from the envelope progressively reduces its density. This allows shorter-wavelength radiation to escape, making the YSO observable in a broader range of wavelengths. Simultaneously, the initially toroidal structure of the envelope changes into a well-defined circumstellar disk. Accretion-driven outflows further sculpt the envelope by creating cavities, providing less obscured sightlines to the central regions. This marks the onset of the Class I stage, where the YSO becomes detectable at MIR wavelengths, extending well below 100~\um. In the subsequent evolutionary phases, namely Class II and III, the envelope continues to dissipate, resulting in a progressively less obscured view of the central protostar and its surrounding disk. During this period, the emission becomes increasingly dominated by radiation from the MIR and NIR bands, originating from the inner regions of the disk and the protostar itself, as the system transits toward the classical pre-Main Sequence phase \citep{Greene1994}.

Combining NIR to sub-millimeter (sub-mm) data provides a comprehensive view of the star formation process.
The \hg\ \citep[Herschel InfraRed Galactic Plane Survey,][]{molinari2010hi} observations aptly trace the very early stages of the star formation process at clump scales, shedding light on the initial steps of the birth of YSOs. \citet{elia2017hi} compiled a catalog of over $10^5$ compact clumps located within the inner Galaxy, categorizing them based on their evolutionary stages. These clumps, with estimated sizes ranging from 0.1 to 1 pc, are substantially larger than the dense cores, typically measured at scales between 0.01 and 0.1 pc \citep{bergin2007cold}. This disparity suggests that these clumps likely harbor clusters of YSOs. However, the limited angular resolution of single-dish telescopes operating in the FIR and sub-mm regimes makes it challenging, if not impossible, to resolve individual YSOs within these clumps. 
To overcome this limitation, FIR/sub-mm imaging must be combined with higher-resolution NIR/MIR observations, which provide a clearer view of individual YSOs. Surveys such as \spitzer\ and \wise\ have been instrumental in this regard.

Over the past few decades, investigations conducted in NIR and MIR spectra further have enhanced our understanding of the interstellar medium (ISM), identifying deeply embedded objects, and distinguishing YSO populations from background field stars. Observatories like the \textit{Spitzer Space Telescope} \citep{werner2004spitzer} and the \textit{Wide-field Infrared Survey Explorer} \citep[WISE,][]{wri10} have been particularly effective in detecting deeply embedded YSOs, providing insights into the later phases of star formation.

To illustrate the advantage of multi-wavelength observations, Figure~\ref{fig:ds9} presents a portion of the Galactic plane observed across different wavelengths, with \hg\ clumps overlaid as ellipses. This visualization helps identify potential YSO counterparts \citep{maruccia,koe14,evans2009} and their spatial distribution within the star forming regions \citep{retter2021spatial, megeath2015spitzer, yadav2016multiwavelength}. Furthermore, this approach facilitates the modeling of the spectral energy distribution (SED) of star-forming clumps across a wide wavelength range, providing critical statistical information to characterize their evolutionary stages \citep[e.g.][]{persi2016complex,tapia2018star}. 

However, the final dataset is often made up of a rich set of heterogeneous and intricate features, and traditional analytical methods could not be able to uncover hidden patterns and relationships within this complex data landscape. 

In this perspective, this work aims at characterizing the evolutionary path of star forming clumps in \hg\ by employing a systematic four-step approach. First, we perform feature engineering to construct and refine a comprehensive dataset of multi-wavelength parameters, characterizing the physical and observational properties of clumps. Second, we apply feature selection techniques to identify the most relevant and discriminative parameters for describing the star formation processes. Finally, we conduct a classification analysis using the full original parameter space and compare it with the reduced space defined by the selected features. 
Our goal is to identify key features, eliminate redundant parameters, and reveal hidden relationships within the dataset. This will enable a more precise and insightful characterization of star formation processes, with the final aims of: (a) obtaining a robust and informative set of features capable of accurately classifying the clumps according to their evolutionary properties, and (b) investigating whether a connection exists between the cold material reservoir within the cold dust component traced by FIR/sub-mm emission and the already formed YSOs, which are the precursors of stars. To this purpose, we design our experiments to assess whether the FIR/sub-mm properties alone are sufficient to predict the evolutionary stage of a clump, without relying on direct information from NIR/MIR tracers of the embedded YSOs.
To the best of our knowledge, this data-driven approach is the first attempt aimed at replacing traditional hand-crafted analyses in this context. This motivated our focus on the exploration of the parameter space to identify the physical features that most effectively discriminate between different evolutionary stages of clumps. In this sense, the pioneering use of data-driven techniques for parameter space exploration on this astrophysical topic has the main merit of inferring possible new and deeper knowledge on the physics of the problem.

\textbf{Outline:}
The paper is structured as follows. Section~\ref{sec:dataset} presents the dataset, followed by Section~\ref{sec:featureEnginereeing}, which details the feature engineering process. In Section~\ref{sec:methods}, we outline the feature selection and classification methods, while Section~\ref{sec:experiments} describes our experiments. Finally, we discuss the results in Section~\ref{sec:discussions} and conclude in Section~\ref{sec:conclusions}.

\section{\hg\ clumps and NIR/MIR: building a multi-faceted dataset}\label{sec:dataset}
In the attempt to better investigate the star formation process, we adopt a collection of catalogs at different wavelengths. In particular, we employ the following surveys, as they provide near-complete coverage from the NIR/MIR to the sub-mm wavelength range, thus sampling the full SED of the clumps and their embedded YSOs:

\begin{itemize}
    \item the \hg~photometric catalog\footnote{This dataset is available in VIALACTEA project, at \url{http://vialactea.iaps.inaf.it/vialactea/public/HiGAL_360_clump_catalogue.tar.gz}, and published in \cite{elia2017hi}.}. Introduced by \cite{mol16a}, this comprehensive catalog of compact sources within the inner regions of the Milky Way, contained in a 2\arcdeg~latitude strip, comprises data gathered from the five Herschel bands at 70, 160, 250, 350, and 500~$\mu$m, each offering distinct perspectives with nominal angular resolutions of 5\arcsec, 12\arcsec, 18\arcsec, 24\arcsec, and 36\arcsec, respectively. However, due to the limited available transmission bandwidth, the PACS 70~$\mu$m and 160~$\mu$m channels were co-added on-board Herschel. The resulting instrumental beams are 5.8\arcsec $\times$ 12.1\arcsec and 11.4\arcsec $\times$ 13.4\arcsec, respectively. Both beams are elongated along the scan direction \citep{mol16a}. Compact sources were extracted using the CuTEx algorithm~\citep{mol11}, specifically designed for detecting sources in a highly complex and non-uniform background.
    For our analysis, we use both the \hg\ fluxes (from 70$~\mu$m to 500~$\mu$m) and the physical parameters of the clumps. For details on how these parameters were derived, we refer the reader to \cite{elia21}. ;
    \item the \gli~I (Galactic Legacy Infrared Mid-Plane Survey Extraordinaire) catalog\footnote{This catalog is available at \url{https://irsa.ipac.caltech.edu/data/SPITZER/GLIMPSE/overview.html}.}~\citep{ben03,chu09}. This survey, conducted in the mid-infrared range, spans a critical longitude range between 10\arcdeg$ <\ell<$ 65\arcdeg, $|b|\leq$ 1, focusing on the inner regions of the Galactic plane. GLIMPSE I captures data across a range of wavelengths, in the 3.6, 4.5, 5.8, and 8.0~$\mu$m bands, by using the IRAC instrument~\citep{faz04};
    \item the AllWISE Source Catalog\footnote{Data can be downloded at \url{https://irsa.ipac.caltech.edu/Missions/wise.html}.}. This survey encompasses a vast array of data obtained from the Wide-field Infrared Survey Explorer (\wise) mission, which mapped the entire sky in four infrared bands: 3.4, 4.6, 12, and 22~$\mu$m (hereinafter, w1, w2, w3, and w4, respectively), boasting angular resolutions of 6.1, 6.4, 6.5, and 12 arcseconds, respectively~\citep{wri10};
    \item the \uk~Galactic Plane Survey (GPS)\footnote{The UKIDSS Eighth Data Release (DR8), including catalogues and images, may be accessed from the WFCAM Science Archive (WSA) at \url{http://wsa.roe.ac.uk/}}. Conducted by the United Kingdom Infrared Deep Sky Survey (\uk) project, the GPS utilizes the United Kingdom Infrared Telescope (UKIRT) to observe the Galactic Plane at 1.25, 1.65, and 2.20~$\mu$m (namely, J, H, and K), covering $\sim1868$ deg$^2$ of the north plane, in the coordinate ranges 15\arcdeg$<\ell<107$\arcdeg and 141\arcdeg$<\ell<230\arcdeg$, with $|b|<5$\arcdeg~\citep{luc08}.
\end{itemize}

\section{Feature Engineering}\label{sec:featureEnginereeing}
\begin{figure}\centering
	\includegraphics[width=.7\columnwidth]{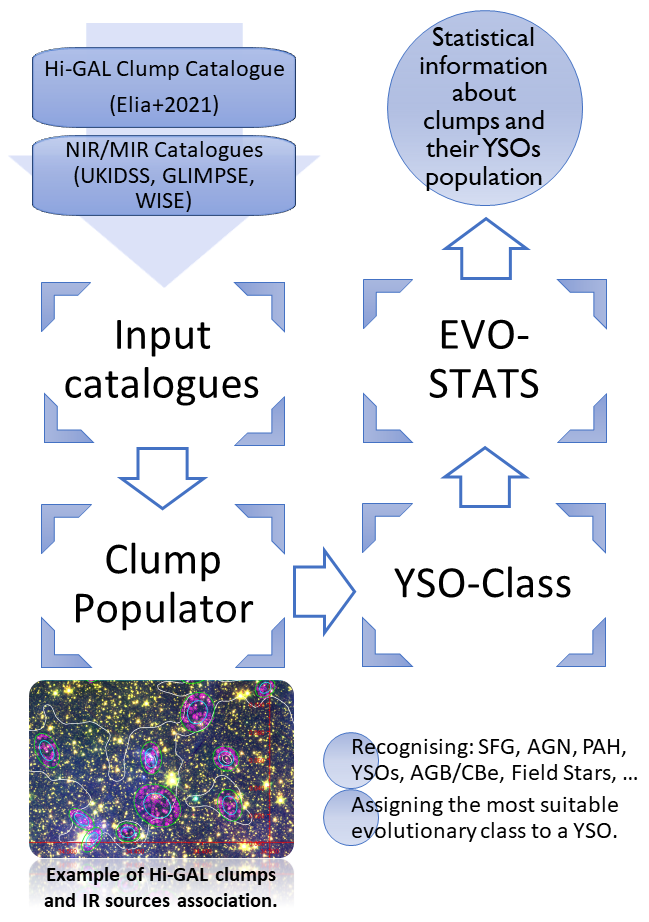}
    \caption{The feature engineering process adopted for obtaining a multi-wavelength catalog of both statistical and physical features of the \hg\ clumps within the Galactic plane region 10\arcdeg$\leq \ell \leq$60\arcdeg, $|b| \leq 1^{\circ}$. For further details, see \citetalias{maruccia}.}
    \label{fig:featureengineeringworkflow}
\end{figure}

We follow the feature engineering workflow described in detail by \citet[][hereinafter M25]{maruccia} and illustrated in Figure~\ref{fig:featureengineeringworkflow}. Our goal is to create a multi-wavelength catalog that statistically and physically describes all clumps within the Galactic plane region 10\arcdeg$\leq \ell \leq$60\arcdeg, $|b| \leq 1^{\circ}$. 
We briefly summarize the whole procedure in the following.

\subsection{\textit{ClumpPopulator}}\label{sec:clumppopulator}
The first module is called \textit{ClumpPopulator}, a specifically tailored version of the C$^3$ tool \citep{ric17}. The \textit{ClumpPopulator} module identifies all IR sources located within \hg\ clumps. Each clump is represented as an elliptical region centered on its Galactic coordinates, with axes defined by the Full Width at Half Maximum (FWHM) of a 2D Gaussian fit to its $350~\mu m$ source profile.
Given the \hg\ clumps, the \textit{ClumpPopulator} module has been run for three times, thus obtaining three lists of associations clumps-IR sources: one for \gli, one for \wise, and one for \uk, respectively. 

\subsection{\STEvMSCARS}\label{sec:stemscars}
The three lists in output from the previous step, have been then positionally cross-matched for obtaining a multi-wavelength sources catalog, which has been given in input of the second module of our evolutionary workflow, called \STEvMSCARS. This component is crucial, as it assigns the most suitable evolutionary class to a source, investigating its color-color or magnitude-colour diagrams through a set of specific prescriptions. These prescriptions finally classify each IR sources, distinguishing whether it belongs to one of the following categories: Star Forming Galaxies (SFGs), Active Galactic Nuclei (AGNs), unresolved knots of shock emission, sources where polycyclic aromatic hydrocarbons (PAH) emission contaminates the photometric apertures, YSOs (Class I/II), AGB/CBe Stars, Field Stars \citep{gut08, gut09, koe14}.

\subsection{\textit{Evo-Stats}}\label{sec:evostats}
A further step after classification, called \textit{Evo-Stats}, provides the statistical information about the YSOs population associated to each \hg~clump. Indeed, in this step the data are aggregated to extract key information about each clump. This includes determining the number and evolutionary stage distribution of YSOs, identifying the detected field stars, quantifying and categorizing contaminants, and assessing the spatial distribution and density of the YSOs within clumps. Furthermore, for each clump a \textit{clump class} is assigned, calculated as the average class of all the YSOs within the clump. This classification provides an intermediate value between the canonical Classes I and II, serving as an indicator of the clump's overall evolutionary stage.

\subsection{Resulting features}
From the evolutionary workflow outlined in Sections~\ref{sec:clumppopulator}--\ref{sec:evostats}, we obtain a final catalog of the \hg~clumps within the Galactic region $10^{\circ} \leq \ell \leq 60^{\circ}$, composed by 6,940 clumps populated by a total number of 10,600 YSOs. In this catalog, each clump is described by a large set of features, which can be divided into photometric and physical features. 

The photometric features refer to the fluxes and their combinations useful to derive the colors characterizing the clump emission. Firstly, it is worth noting that, since each clump may host multiple YSOs emitting in the NIR/MIR range, we need to characterize these clumps by aggregating the properties of the individual YSOs within them. To this aim, fluxes and magnitudes of the individual YSOs are extracted from the respective catalogs presented in Section~\ref{sec:dataset}. 
Unlike previous studies, if a YSO is not detected at some bands, we manage these missing values adopting the sensitivity limits provided by the respective surveys. Adopting these limits in cases of non-detection enables us to retain and maximize the use of available information, rather than discarding potentially valuable data. The adopted sensitivity limits in magnitude (corresponding to upper limits in flux) are: 15.5, 15.0, 13.0, and 13.0 for the \gli\ bands at 3.6, 4.5, 5.8, and 8.0~$\mu$m \citep{chu09}; 16.9, 16.0, 11.5, and 8.0 for the \wise\ bands w1 through w4 \citep{cutri2013explanatory}; and 19.8, 19.0, and 18.1 for the \uk\ J, H, and K bands, respectively \citep{ukidss_mag}. 
Moreover, the Galactic plane in the range $10\arcdeg \leq \ell \leq 60\arcdeg$ is heavily affected by interstellar extinction. Therefore, to reliably apply color–magnitude diagnostics to stellar sources in this region, it is necessary to correct their observed fluxes for dust extinction along the line of sight. We adopt the extinction laws from \citet{rieke1985interstellar} for the $JHK$ bands and from \citet{xue2016precise} for the MIR bands (\gli\ and \wise). Additionally, we assume an average extinction of 1A$_\mathrm{V}$ per kiloparsec of heliocentric distance \citep{juvela2016allsky}. Since these YSOs belong to clumps, the clump distances, available in the \hg\ catalog, are used to estimate the extinction correction. More details about these steps can be found in \citetalias{maruccia}.
Starting from these data, we compute the following photometric features:
the total flux for each band, calculated as the sum (hereinafter referred to as ``sum'') of the fluxes of all YSOs within the same clump; the median (referred to as ``med'') of the YSO fluxes within the clump; the absolute value of the difference between the average and, respectively, the minimum and the maximum of the YSOs fluxes within the clump (referred to as ``D1'' and ``D2'', respectively). As regards the colors, we combine the magnitudes of each individual YSO, obtaining six colors for \gli\ and \wise\, and three for \uk. Subsequently, we aggregate these values for each clump, taking into account the minimum value (``min''), the maximum (``max''), the median (``med''), and, as done for the fluxes, the absolute value of the difference between the average and, respectively, the minimum and the maximum of the given color (``D1'' and ``D2''). To provide a more comprehensive description of the nature of clumps in the MIR/FIR range, we utilize the clump fluxes at the wavelengths investigated by Herschel (as described in Section~\ref{sec:dataset}), along with the following colors: $[w4-70]$, which is a combination of the \wise\ flux at $22~\mu$m and the Herschel 70~$\mu$m, $[70-250]$, and $[250-350]$. Finally, all fluxes have been rescaled to a reference distance of 1~kpc using the known distances of clumps.

The physical features mainly refer to the \hg~clumps properties presented in the \cite{elia21}'s catalog. In particular, they include the bolometric luminosity ($L_{\textrm{bol}}$), the mass (\textit{M}), the dust temperature ($T_{\textrm{dust}}$), the bolometric temperature ($T_{\textrm{bol}}$), the surface density (\textit{$\Sigma$}), the \lm\ ratio, the ratio between the bolometric luminosity and its fraction computed over the range $\lambda \geq 350~\mu$m ($L_{\textrm{ratio}}$). We refer to the original paper for further details on how these physical properties were calculated.

Finally, we consider the YSO density ($\delta_{YSO}$), a feature generated by the \STEvMSCARS\ procedure, and calculated as the fraction of the YSOs among the total number of sources falling within the same clump. Table~\ref{tab:overview_res} summarizes all the features that we want to investigate in our work through experiments of feature selection and classification.

\begin{table}[!htpb]
    \centering
    \resizebox{\columnwidth}{!}{  
    \Huge
    \begin{tabular}{ccccc}
\hline
\hline
&\textbf{Source} & \textbf{Features} & \textbf{Base PS} & \textbf{Enriched PS} \\\hline
 \multirow{6}{*}{\rotatebox[]{90}{\textbf{fluxes}}}&\gli & \multirow{5}{*}{\Centerstack[c]{sum, med\\ D1, D2\\ of individual\\fluxes}} &---& D1\_i4, sum\_i4 \\\cdashline{2-2}\cdashline{4-5}

&\wise & &---& \Centerstack[c]{D2\_w1, sum\_w1, med\_w1,\\ sum\_w2, med\_w2, D1\_w3,\\ D1\_w4, sum\_w4, med\_w4} \\\cdashline{2-2}\cdashline{4-5}

&\uk & &---& D2\_J, med\_J, D2\_H \\\cdashline{2-5}

&\hg & \Centerstack[c]{$F_{\textrm{70}}$, $F_{\textrm{160}}$, $F_{\textrm{250}}$,\\$F_{\textrm{350}}$, $F_{\textrm{500}}$} & $F_{\textrm{160}}$, $F_{\textrm{500}}$ & --- \\ \hline 

\multirow{13}{*}{\rotatebox[]{90}{\textbf{colors}}}&\gli & \multirow{8}{*}{\Centerstack[c]{min, max, med\\ D1, D2\\ of individual\\colors}} &---& {\Centerstack[c]{D2\_i1-i2, min\_i1-i2, \\med\_i1-i2, max\_i1-i2, \\med i1-i3, max\_i1-13, \\med\_i1-i4, min\_i2-13, \\med\_i2-13, max\_i2-13, \\min\_i2-i4, med\_i2-i4, \\max\_i2-i4, min\_i3-i4, \\med\_i3-i4, max\_i3-i4}} \\\cdashline{2-2}\cdashline{4-5}

&\wise & &---& {\Centerstack[c]{min\_w1-w2, med\_w1-w2, \\max\_w1-w2, min\_w1-w3, \\med\_w1-w3, max\_w1-w3,\\min\_w2-w3, med\_w2-w3, \\max\_w2-w3}} \\\cdashline{2-2}\cdashline{4-5}

&\uk & &---&\makecell{ min\_H-K, med\_H-K,\\ max\_H-K} \\ \cdashline{2-5}

&\hg & {\Centerstack[c]{[70-250],\\[250-350]}} & [250-350] & {\Centerstack[c]{---}} \\\cdashline{2-5}

&\wise/\hg & [w4-70] & [w4-70] & --- \\ \hline 

\multirow{2}{*}{\rotatebox[]{90}{{\textbf{physical}}}}&\hg & {\Centerstack[c]{\textit{M}, $L_{\textrm{ratio}}$,\\ \lm\,$L_{\textrm{bol}}$, \\$T_{\textrm{dust}}$, $T_{\textrm{bol}}$,\\ \textit{$\Sigma$}}} & {\Centerstack[c]{$L_{\textrm{ratio}}$,\\$T_{\textrm{bol}}$,\\ \textit{$\Sigma$} }} & {\Centerstack[c]{---}} \\\cdashline{2-5}

&\STEvMSCARS & $\delta_{YSO}$ & $\delta_{YSO}$ & $\delta_{YSO}$ \\ \hline 
    \end{tabular}
    }
    \caption{Overview of all the features adopted for and obtained from feature selection in our experiments. The first half of the table (fluxes and colors) is related to the photometric features, while the second half corresponds to the physical features. See the text for further details.}

    \label{tab:overview_res}
\end{table}

\section{Workflow}\label{sec:methods}
The workflow adopted in this paper involves a structured, multi-step approach to ensure the identification of the most relevant features involved in the star formation process and evaluate the results with a classification step. 
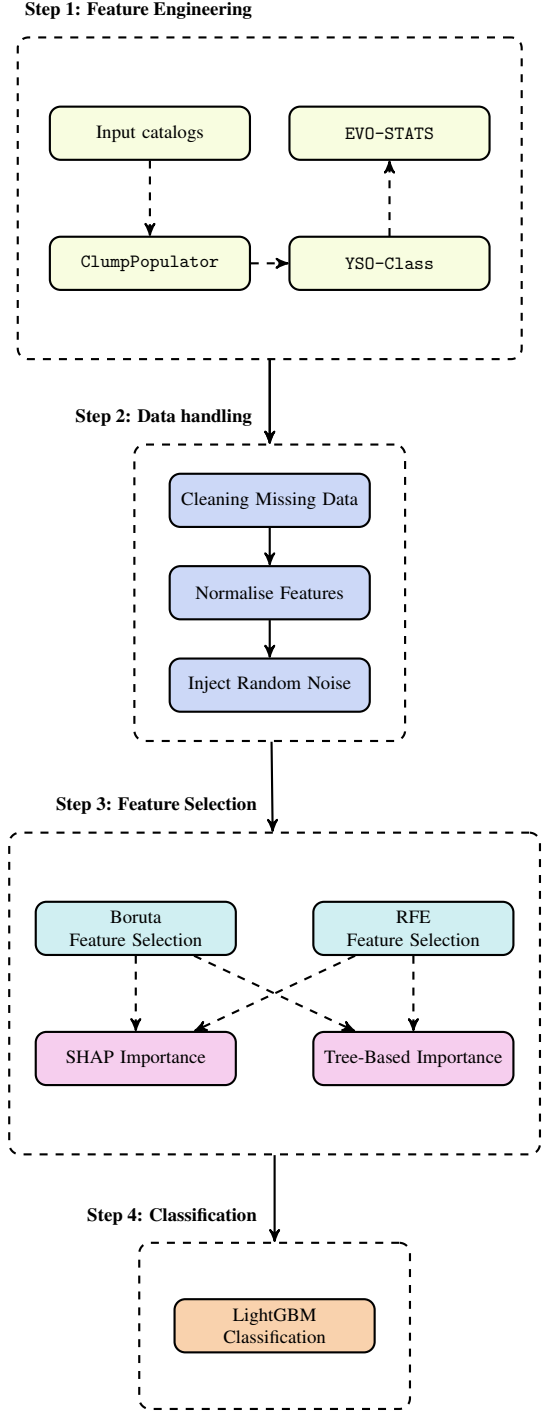
\begin{figure}[!hbpt]
    \centering
\usetikzlibrary{shapes,arrows,positioning,decorations.pathreplacing}
\begin{tikzpicture}[>=stealth',auto,node distance=2cm, 
yellowrectangle/.style={rectangle, draw=black, fill=yellowii!20, rounded corners, align=center, minimum width = 3.5cm, minimum height = 1cm, text width=3.5cm}, 
bluerectangle/.style={rectangle, draw=black, fill=bluev!20, rounded corners, align=center, minimum width = 3.5cm, minimum height = 1cm, text width=3.5cm}, 
greenrectangle/.style={rectangle, draw=black, fill=greenii!20, rounded corners, align=center, minimum width = 3.5cm, minimum height = 1cm, text width=3.5cm}, 
pinkrectangle/.style={rectangle, draw=black, fill=pinki!20, rounded corners, align=center, minimum width = 3.5cm, minimum height = 1cm, text width=3.5cm},
orangerectangle/.style={rectangle, draw=black, fill=orangei!40, rounded corners, align=center, minimum width = 3.5cm, minimum height = 1cm, text width=3.5cm},
steptext/.style={text width=5cm, align=left},
dashedbox/.style={draw=black, dashed, rounded corners, inner sep=1.2cm},
dashedbox2/.style={draw=black, dashed, rounded corners, inner sep=1.8cm},
thick,scale=0.705, every node/.style={scale=0.705}
]

\node [yellowrectangle] (input_cat) {Input catalogs};
\node [yellowrectangle, below=1cm of input_cat] (clump_pop) {\cp};
\node [yellowrectangle, right=0.5cm of clump_pop] (YsoClass) {\STEvMSCARS};
\node [yellowrectangle, right=0.5cm of input_cat] (EvoStats) {\evo};

\node [dashedbox2, fit={(input_cat) (clump_pop) (YsoClass) (EvoStats)}] (step1box) {};

\node [above left=0.2cm of step1box.north, font=\bfseries] {Step 1: Feature Engineering};

\node [bluerectangle, below=1.5cm of step1box.south] (preprocess) {Cleaning Missing Data};
\node [bluerectangle, below=0.5cm of preprocess] (normalize) {Normalise Features};
\node [bluerectangle, below=0.5cm of normalize] (injectnoise) {Inject Random Noise};

\node [dashedbox, fit={(preprocess) (normalize) (injectnoise)}] (step2box) {};

\node [above left=0.2cm of step2box.north, font=\bfseries] {Step 2: Data handling};

\node [greenrectangle, below=2.5cm of injectnoise, xshift=-2.5cm] (boruta) {Boruta\\Feature Selection};
\node [greenrectangle, right=1.cm of boruta] (rfe) {RFE\\Feature Selection};
\node [pinkrectangle, below=1cm of rfe] (treebased) {Tree-Based Importance};
\node [pinkrectangle, below=1cm of boruta] (shap) {SHAP Importance};

\node [dashedbox2, fit={(rfe) (boruta) (treebased) (shap)}] (step3box) {};

\node [above left=0.2cm of step3box.north, font=\bfseries] {Step 3: Feature Selection};

\node [orangerectangle, below=1.9cm of step3box.south] (LGBM) {LightGBM\\Classification};

\node [dashedbox, fit={(LGBM)}] (step4box) {};

\node [above left=0.2cm of step4box.north, font=\bfseries] {Step 4: Classification};

\draw [->, draw=black] (step1box) -- (step2box);
\draw [->, draw=black] (preprocess) -- (normalize);
\draw [->, draw=black] (normalize) -- (injectnoise);
\draw [->, draw=black] (step1box) -- (step2box);
\draw [->, draw=black] (step2box) -- (step3box);
\draw [->, draw=black] (step3box) -- (step4box);

\draw [->, draw=black, dashed] (boruta) -- (shap);
\draw [->, draw=black, dashed] (boruta) -- (treebased);

\draw [->, draw=black, dashed] (rfe) -- (shap);
\draw [->, draw=black, dashed] (rfe) -- (treebased);

\draw [->, draw=black, dashed] (input_cat) -- (clump_pop);
\draw [->, draw=black, dashed] (clump_pop) -- (YsoClass);
\draw [->, draw=black, dashed] (YsoClass) -- (EvoStats);

\end{tikzpicture}

    \caption{The four-step workflow adopted in this work. Firstly, the feature engineering step builds the parameter space from several input catalogs (\textbf{Step 1: Feature Engineering}). More details about this step can be found in Section~\ref{sec:featureEnginereeing}, while the feature engineering process is depicted in Figure~\ref{fig:featureengineeringworkflow}. After cleaning and normalizing the dataset (\textbf{Step 2: Data Handling}), the most relevant features are identified (\textbf{Step 3: Feature Selection}) and used for classification tests (\textbf{Step 4: Classification}).}
    \label{fig:fs_procedure}
\end{figure}
This process, depicted in Figure~\ref{fig:fs_procedure}, is made up of the following steps:
\begin{itemize}
    \item Step 1: Feature Engineering. It has already been discussed in detail in Section~\ref{sec:featureEnginereeing} and represented in Figure~\ref{fig:featureengineeringworkflow}. This step involves the construction of a parameter space (hereinafter, PS) that encapsulates the key aspects of the star formation process. Utilizing the input catalogs as a foundation, we systematically derived a comprehensive set of features essential for describing and analyzing the phenomenon under investigation.
    \item Step 2: Data handling. It consists in preparing the dataset for the next stages of analysis, in particular cleaning the missing data, normalizing the features to ensure equal contribution of each feature and prevent scale-related biases, and injecting random noise to improve model robustness, prevent overfitting, and increase diversity in the parameter space. More details are provided in Section~\ref{sec:preliminarities}.
    \item Step 3: Feature Selection. In this stage different algorithms have been used for identifying the most relevant features in the PS, as well described in Section~\ref{sec:featureselection}. Results are shown in Section~\ref{sec:experiments_FS}.
    \item Step 4: Classification. The final step involves classification using the Light Gradient Boosting Machine (hereinafter, LightGBM) to compare results across different PS. LightGBM’s leaf-wise growth and histogram-based algorithm improve accuracy and efficiency over traditional GBDT methods. More details are provided in Section~\ref{sec:lgbm}. Results are discussed in Section~\ref{sec:experiments_classification}.
\end{itemize}

\subsection{Data handling}\label{sec:preliminarities}
The second step of the workflow consists in managing our dataset, to ensure consistency and mitigate biases. This phase encompasses three processes aimed at preparing data for subsequent analysis: cleaning missing data, feature normalization, and the injection of random noise. The first process involves cleaning missing data, replacing missing values with statistical measures or removing incomplete entries, generating new classes, and removing irrelevant features. These preliminary actions are essential for ensuring that the dataset is both accurate and relevant. Normalizing the data is an important step \citep{huang2023normalization}, particularly when the features have different scales. Techniques like Min-Max normalization or Z-score standardization are the most popular choice for normalizing data \citep{singh2022feature} and ensure that each feature contributes equally to the model by adjusting the range of the data. This helps avoid problems like certain features dominating the model due to their larger scale or dynamics. Furthermore, random noise can be injected into the data \citep{grandvalet1997noise, hua2006noise}, introducing artificial features that help improve the robustness of the model by preventing it from overfitting to the noise present in real-world data. This technique can also help increase the diversity of the PS, making the model more generalizable. Once the data has been cleaned and normalized, the dataset is ready to be used in the next steps of our analysis.

\subsection{Feature Selection}\label{sec:featureselection}
Finally, in the third step of our workflow we focus on feature selection (hereinafter, FS), a critical phase for identifying the most relevant features in our PS. In this phase, we adopt two different algorithms for the FS. In particular:
\begin{itemize}
    \item the Recursive Feature Elimination (RFE), which evaluates recursively the importance of a  feature by removing one feature at a time from PS and calculating its impact on classification prediction. Then, it selects those features (columns) in a training dataset that are more or the most relevant in predicting the target variable.
    \item the Boruta algorithm \cite{boruta}, which identifies the most relevant features in a dataset by comparing their importance to randomly generated ``shadow features''. It begins by creating a copy of the original dataset's features, where the values in each column are shuffled to eliminate any inherent relationship with the target variable. These shuffled features, known as shadow features, are then merged with the original dataset to form a new feature space  whose dimension is twice the original dataset. At this point, the Random Forest \citep{breiman2001random} works on these extended features space and determines their importance \citep{kursa2011all} using a statistical test, often the Z-Score, which quantifies the contribution of a feature to the model’s performance. The algorithm checks if each original feature has a higher importance score than the maximum importance of the shadow features. If an original feature has a higher importance score than all shadow features, the feature is considered significant and kept in the dataset. Otherwise, features having a lower score than the shadow features are considered insignificant and dropped from the dataset. Features whose importance cannot be definitively determined are labeled as tentative and re-evaluated in subsequent iterations. In each iteration, the Boruta algorithm uses only the retained and tentative features to create a new set of shadow features, repeating the process of evaluating their importance relative to noise. This iterative approach ensures that features are either confirmed as significant or discarded as irrelevant. This is repeated until a specified number of iterations has been achieved or when all features have been confirmed or dropped. It is worth noting that, unlike traditional FS techniques, the Boruta algorithm is tasked with finding a minimal optimal feature set rather than finding all the features relevant to the target variable. By using shadow features as a baseline, it offers a stringent and interpretable mechanism to filter out noise and irrelevant variables, making it especially effective in high-dimensional datasets where irrelevant features could obscure meaningful patterns.
\end{itemize}
Furthermore, each FS algorithm is paired with two methods for calculating the feature importance:
\begin{itemize}
    \item the standard tree-based importance, which evaluates the contribution of each feature considering the standard tree-based feature importance calculation. This evaluation is typically based on metrics like the Gini importance or mean decrease in impurity \citep{breiman2017classification}, which measure how much a feature reduces uncertainty or improves model performance at decision splits across the trees in the ensemble. Features are ranked according to the magnitude of their importance scores, and the less important features (i.e. those with lower scores) are considered redundant or noisy as they contribute minimally to reducing prediction error, thus they are excluded from the final selection. This approach not only simplifies the model by reducing the number of input features but also helps prevent overfitting, as irrelevant features can introduce noise and decrease the model's generalization capability.
    \item the SHAP \citep[SHapley Additive exPlanations,][]{lundberg2017unified} importance, which leverages the Shapley’s values, a concept rooted in game theory, to calculate the contribution of each feature to a model's predictions. This approach determines a feature's importance by comparing the model's predictions with and without the inclusion of that feature, effectively measuring how much the feature influences the output. Inspired by game theory, which evaluates the role of individual players in a collaborative team, SHAP adapts this framework to assess features within a machine learning model. In game theory, a player can either participate in or abstain from a game, and SHAP translates this idea by defining a feature as ``joining the model'' when its value is known and ``not joining the model'' when its value is missing or replaced. The calculation involves considering all possible combinations of features, assessing how the model's output changes when the feature of interest is added or removed in each scenario. This ensures that SHAP provides a fair and consistent estimate of feature importance, accounting for the interactions between features. The result is a detailed, instance-specific measure of importance that not only identifies which features are influential but also quantifies their impact in a way that is interpretable and intuitive, making SHAP a powerful tool for explaining complex models.
\end{itemize}
This results in four main combinations (RFE + standard tree-based importance, RFE + SHAP importance, Boruta + standard tree-based importance, Boruta + SHAP importance), and we evaluate the each outcomes to determine which combination performs best for our PS. The selected features from this step will then be used in the subsequent classification stage.

\subsection{LightGBM Classification}\label{sec:lgbm}
We now focus on the classification, with the primary goal of comparing results across the different PS identified in the FS phase. For this purpose, we employed LightGBM, an advanced implementation of the Gradient Boosting Decision Tree (GBDT) algorithm \citep{ke2017lightgbm}. LightGBM avoids the typical hierarchical growth approach used by most GBDT algorithms and instead employs a leaf growth algorithm with depth constraints. In LightGBM, the tree grows leaf-by-leaf, where the leaf with the highest split gain among all current leaves is selected and split, and this process continues iteratively. This leaf-by-leaf growth method tends to reduce errors and achieve better accuracy with the same number of splits compared to the traditional layer-by-layer approach. The histogram for the next layer node is calculated by finding the difference between its parent and sibling nodes. The histogram is computed by traversing the leaf nodes according to the growth strategy. LightGBM uses a histogram-based algorithm, which has the advantage of efficiently calculating the histogram of a leaf node while simultaneously obtaining the histogram of its sibling node at half the computational cost, making the process twice as fast \citep{ke2017lightgbm}.

\section{Experiments}\label{sec:experiments}
As described in the previous sections, all the considered features are traditionally essential for comprehensively describing the star formation process. They encompass various evolutionary phases observable across a broad spectrum of wavelengths. FS serves as a crucial step in streamlining our analysis. By identifying and eliminating redundant features, we can focus solely on those most pertinent to characterize the star formation process effectively.

We designed two experiments based on different PS. In the first one, referred to as the \textit{base PS}, we include fluxes and colors of the \hg\ clumps, along with their physical properties. The second experiment builds upon this feature set by incorporating the additional photometric features described in Section \ref{sec:featureEnginereeing}. Hereinafter, we will refer to this PS as \textit{enriched PS}. 
For studying the clump evolution, we take into account the {\it clump class} as target, defined as the average of the classes of all the YSOs associated with a given clump, as described in Section~\ref{sec:evostats}.
This setup serves a dual purpose. First, it allows us to obtain a robust and accurate classification of the \hg\ clumps in terms of their evolutionary properties. Second, it enables us to assess whether there is a link between the cold material traced in the FIR/sub-mm (associated with the clumps themselves) and the stellar content already formed within them, observable in the NIR/MIR.
Moreover, since the clump class is derived through a process that involves the additional features present only in the \textit{enriched PS}, our experimental design is also aimed at testing whether the \textit{base PS} alone — i.e., without direct information about the embedded YSOs — is sufficient to predict the clump evolutionary stage. This provides insights into the predictive power of the clump's intrinsic properties and their observed fluxes/colors.

\begin{figure}[!hbpt]
    \centering
	\includegraphics[width=.85\columnwidth]{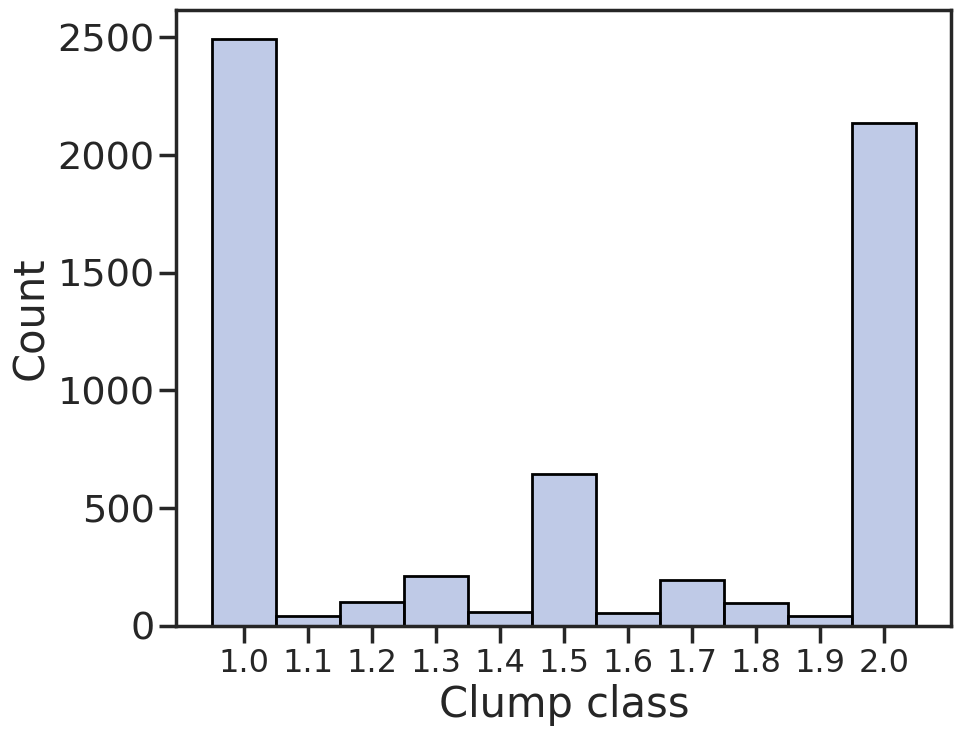}
    \caption{Clump class distribution.}
    \label{fig:clumpclass}
\end{figure}

The starting dataset is made up of 6,079 objects. During the preprocessing step, four clumps have been discarded as they presented \textit{NaN} values in correspondence of some physical features, thus the final clump dataset is made up of 6,075 objects. Before going on with our analysis, the dataset has been normalized using the Z score. 
In Figure~\ref{fig:clumpclass} it is shown the distribution of the clump classes for our dataset. Clumps with Class 1.0 and 2.0 are well distributed in the dataset, and are more numerous than clumps with intermediate classes. 
For our experiments, we decide to label as Class~0 all clumps with class $\le 1.5$, as Class~1 all clumps with class $> 1.5$, resulting, respectively, 3,552 (58\%) and 2523 (42\%) clumps. Finally, in all the following experiments the train+validation/test split was always based on, respectively, 70\% training and validation (the latter amounting to 10\%) and 30\% test of the dataset.

\subsection{Feature Selection}\label{sec:experiments_FS}
FS is an important aspect of our workflow. It aims at simplifying complex datasets into a more manageable form without losing essential information, and selecting only those features that are more involved in the description of the star formation process. Since we are looking to all the relevant features this should bring also useful insight on the physics driving the selection.
We perform the FS on the two different PS, the \textit{base PS} made up of 16 features and the \textit{enriched PS}, made up of 135 features. Moreover, as described in Section~\ref{sec:featureselection}, we combined both the RFE and the Boruta algorithm with two measures of importance, the standard tree-based and the SHAP importances. 
In Table~\ref{tab:expFS} results of these four experiments are summarized. In the experiments with the standard tree-based importance, some of the shadow features have been selected as good features, thus invalidating the final results. The best solution is that one in which we adopted the combination of Boruta and the SHAP importance, both for the \textit{base PS} and the \textit{enriched PS}.

\begin{table}[!hbpt]
    \centering
    \begin{tabular}{lcc}
        \hline
        \hline
        Experiment              & \textit{base PS} & \textit{enrichedPS} \\
        \hline
        All features       & 16    & 135\\
        \hline
        RFE + Std tree-based    & 15    & 102\\
        RFE + SHAP              & 14    & 98\\
        Boruta + Std tree-based & 4     & 28\\
        \rowcolor{lavender}
        Boruta + SHAP           & 8     & 43\\
        \hline

    \end{tabular}
    \caption{Results of the FS on both the two PS, using the combination of RFE + standard tree-based importance, RFE + SHAP importance, Boruta + standard tree-based importance, Boruta + SHAP importance. The numbers indicate the total number of features selected for each experiment, while the row \textit{All features} indicates the initial number of the features in each PS. The colored row indicates the best solution obtained.}
    \label{tab:expFS}
\end{table}

As it can be seen, when considering the \textit{base PS}, the algorithm selected 8 features, while for the \textit{enriched PS} the algorithm selected 43 features. In the third and fourth columns of Table~\ref{tab:overview_res}, the selected features are indicated. In the experiment with the \textit{enriched PS} all the selected features are related to the NIR/MIR properties, which are effectively used in the workflow that computes the \textit{clump class} starting from the color-color diagrams. Only one selected feature, the $\delta_{YSO}$, is not related to the MIR properties. 
On the contrary, in the experiment with the \textit{base PS} (without the MIR properties), all the selected features are related to the clump properties. 
Also in this case the $\delta_{YSO}$ is selected. Finally, only the color [w4-70] contains the information from the MIR. %
Further details about the role of these indicators in understanding the evolution stage of clumps are discussed in Section~\ref{sec:discussions}.

\subsection{Classification}\label{sec:experiments_classification}
As explained in Section~\ref{sec:lgbm}, the results obtained previously by the FS phase are finally used in the classification step. In particular, we run the LightGBM for obtaining a classification in four cases: 
\begin{itemize}
    \item \textit{base PS} and all the 16 features;
    \item \textit{base PS} and the selected 8 features;
    \item \textit{enriched PS} and all the 135 features;
    \item \textit{enriched PS} and the selected 43 features.
\end{itemize}

The hyperparameters used for these experiments are reported in Table~\ref{tab:exp_setup}.
\begin{table}[htbp]
\centering
\resizebox{\columnwidth}{!}{%
\begin{tabular}{lcccc}
\hline
\hline
LightGBM and  & Base PS & Base PS & Enriched PS & Enriched PS \\
experiment setup & All features & Sel. Features & All features & Sel. Features \\
\hline
PS dimension & 16 & 8 & 135 & 43 \\
FS method & - & Boruta+SHAP & - & Boruta+SHAP \\
Learning rate & 0.2 & 0.2 & 0.01 & 0.2 \\
\textit{N\_estimators}* & 800 & 800 & 800 & 800 \\
\textit{Num\_leaves}** & 25 & 40 & 35 & 35 \\
Validation set score*** & 0.605 & 0.606 & 0.122 & 0.115 \\
Blind test set score & 0.655 & 0.656 & 0.953 & 0.953 \\
\hline
\end{tabular}%
}
\\[1ex]
\raggedright
\scriptsize
*Number of boosting rounds. \\
**Maximum number of leaves in one tree. \\
***Validation score computed on a held-out validation set.
\caption{LightGBM and experiment setup} \label{tab:exp_setup}
\end{table}
They were selected using a standard grid search optimization strategy. All remaining LGBM hyperparameters were kept at their default values\footnote{\url{https://lightgbm.readthedocs.io/en/latest/pythonapi/lightgbm.LGBMClassifier.html}} as provided by the official implementation.

\begin{figure}[!hbpt]
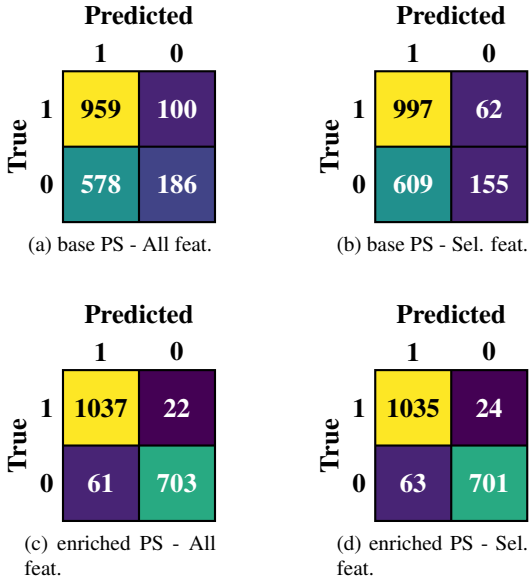

    \centering
    \subfloat[base PS - All feat.]{\hspace{-2.0cm}\ConfusionMatrixViridis{959}{100}{578}{186}\label{fig:onlyclump_all}}
    \hspace{1.5cm}
    \subfloat[base PS - Sel. feat.]{\hspace{-2.0cm}\ConfusionMatrixViridis{997}{62}{609}{155}\label{fig:onlyclump_sel}}\\
    \vspace{0.1cm}
    \subfloat[enriched PS - All feat.]{\hspace{-2.0cm}\ConfusionMatrixViridis{1037}{22}{61}{703}\label{fig:full_all}}
    \hspace{1.5cm}
    \subfloat[enriched PS - Sel. feat.]{\hspace{-2.0cm}\ConfusionMatrixViridis{1035}{24}{63}{701}\label{fig:full_sel}}

    \caption{Confusion matrix for the four experiment of classification adopting the LightGBM. Top panel: \textit{base PS} with all the features (on the left) and the selected features (on the right). Bottom panel: \textit{enrichedPS} with all the features (on the left) and the selected features (on the right).}
    \label{fig:classification_cc}
\end{figure}

In Figure~\ref{fig:classification_cc} we present the results of the four classification experiments, reporting for each case the corresponding confusion matrix. In particular, the top panels illustrate the outcomes obtained when adopting the \textit{base PS} (Figure~\ref{fig:onlyclump_all}), made up of 16 features, alongside the results obtained using the corresponding 8 features (Figure~\ref{fig:onlyclump_sel}) selected during the FS step. As expected, when the classification is performed on the PS with only the clump properties, the experiment suffers the missing information about the NIR/MIR photometric properties. In fact, there is high contamination, especially for the Class 0. This limitation suggests that the information contained in the clump-only PS is insufficient to effectively distinguish between the classes. It is worth noting that, as well explained in Section~\ref{sec:featureEnginereeing}, the clump class is a parameter derived through a process that involves the additional features present only in the \textit{enriched PS,} not in the \textit{base PS}. Analyzing the experimental results from this perspective, we conclude that the clumps intrinsic properties and their fluxes and colors do not have a good predictive power of their evolutionary stages derived from the information about the embedded YSOs. Note that, when the classification is performed using only the 8 selected features, the results do not get worse, as is evident from the upper panels in Figure~\ref{fig:classification_cc} and confirmed by the metrics reported in Table~\ref{tab:exp_metrics} (panels a and b). Thus, the \textit{base PS} appears insufficient to give proper information about the clump class. Nevertheless, the FS step successfully identified the most informative features within the \textit{base PS}, allowing the model to maintain its performance despite the reduced dimensionality.

\begin{table}[!hbpt]
	\centering
	\begin{tabular}{rcccc} 
		\hline\hline
		 & precision & recall & f1-score & support\\
		\hline
		0            & 0.62 & 0.91 & 0.74 & 1,059\\
		1            & 0.65 & 0.24 & 0.35 & 764  \\
                     &      &      &      &      \\
		accuracy     &      &      & 0.63 & 1,823\\
		macro avg    & 0.64 & 0.57 & 0.55 & 1,823\\
        weighted avg & 0.64 & 0.63 & 0.58 & 1,823\\
		\hline\\\multicolumn{5}{c}{(a) \textit{base PS} - All feat.}\\  \\

		\hline\hline
		 & precision & recall & f1-score & support\\
		\hline
		0            & 0.62 & 0.94 & 0.75 & 1,059\\
		1            & 0.71 & 0.20 & 0.32 & 764  \\
                     &      &      &      &      \\
		accuracy     &      &      & 0.63 & 1,823\\
		macro avg    & 0.67 & 0.57 & 0.53 & 1,823\\
        weighted avg & 0.66 & 0.63 & 0.57 & 1,823\\
		\hline\\\multicolumn{5}{c}{(b) \textit{base PS} - Sel. feat.}\\  \\

		\hline\hline
		 & precision & recall & f1-score & support\\
		\hline
		0            & 0.94 & 0.98 & 0.96 & 1,059\\
		1            & 0.97 & 0.92 & 0.94 & 764  \\
                     &      &      &      &      \\
		accuracy     &      &      & 0.95 & 1,823\\
		macro avg    & 0.96 & 0.95 & 0.95 & 1,823\\
        weighted avg & 0.96 & 0.95 & 0.95 & 1,823\\
		\hline\\\multicolumn{5}{c}{(c) \textit{enrichedPS} - All feat.}\\  \\
        

		\hline\hline
		 & precision & recall & f1-score & support\\
		\hline
		0            & 0.94 & 0.98 & 0.96 & 1,059\\
		1            & 0.97 & 0.92 & 0.94 & 764  \\
                     &      &      &      &      \\
		accuracy     &      &      & 0.95 & 1,823\\
		macro avg    & 0.95 & 0.95 & 0.95 & 1,823\\
        weighted avg & 0.95 & 0.95 & 0.95 & 1,823\\
		\hline\\\multicolumn{5}{c}{(d) \textit{enrichedPS} - Sel. feat.}
	\end{tabular}
    \caption{Performance evaluation of the classification step across the four experiments, presenting: \textit{precision} (the accuracy of positive predictions), \textit{recall} (the model's ability to find all positive instances), \textit{F1-score} (a balanced measure of precision and recall), \textit{support} (the number of samples in each class), \textit{accuracy} (the percentage of correctly classified samples), \textit{macro avg} (the average of precision, recall, and F1-score across classes without considering class imbalance), and \textit{weighted avg} (the average of precision, recall, and F1-score across classes, weighted by the support of each class).}
    \label{tab:exp_metrics}
\end{table}

When we consider the \textit{enriched PS}, the results improve substantially compared to the previous experiment. This is evident in the confusion matrices shown in Figure~\ref{fig:full_all} (with all the features) and Figure~\ref{fig:full_sel} (with the 43 selected features). In both cases, there is a very low contamination between the two classes, underscoring the superior predictive power of the full parameter set.
Furthermore, the metrics reported in the panel c and d of Table~\ref{tab:exp_metrics} confirm that the classification performance remains consistent when transitioning from the \textit{enriched PS} to the subset of selected features. These results demonstrate that the \textit{enriched PS} is more effective in capturing the characteristics required to accurately classify clumps, probably due to the fact that the clump class depends on the information about the embedded YSOs. Moreover, the selected features still appear a good choice for modeling the problem, ensuring good results in predicting the right \textit{clump class}. This highlights the robustness of the FS process, which ensures that the selected features retain the critical information necessary for modeling the problem effectively while reducing computational complexity.

\section{Discussion}\label{sec:discussions}
We performed ten different runs for each kind of experiment, both for classification and feature selection use cases, by randomly shuffling the training/test data selection, in order to keep under control the performance of models with respect to the potential variability of the parameter space coverage induced by the different choice of training and blind test samples, as well as to prevent possible selection criteria due to a particular model setup.

In the specific case of classification, this mechanism is always coupled with the model hyperparameter optimization, based on grid search, ensuring the optimal performance at each repeated experiment run. As the final choice of the best result, we always adopt the ``worst case'' strategy, by considering as final result the worst statistical performance in terms of confusion matrix, in order to remain as much as possible conservative regarding the final classification trade-off between the class purity and completeness.

In the case of feature selection, the strategy of repeating ten times the experiment was particularly useful for Boruta method, in order to keep under control the potential fluctuation of the measured relevance of features, induced by the random mechanism of shadow feature construction. As expected, the stability of the model, in principle declared by \cite{boruta}, was confirmed in our experiment, in which always the same features were labeled as true (rank 1, i.e. selected) and false (rank 4, i.e. rejected), without any presence of ``tentative'' features, which can be interpreted as a high reliability of the selected features with respect to the random noise injected into the parameter space and a high robustness with respect to the feature distributions within the parameter space.

The impact of the repeated experiments on the feature importance was also useful to evaluate their potential fluctuation in the case of standard Z score used by the RFE method. Also in this case no any significant variation was detected in terms of selected features, partially due to the LightGBM, which is known to be less sensible to randomness than the classical Random Forest when a hyperparameter optimization, based on the grid search, is performed (which also helps to prevent the occurrence of overfitting, which boosting models may typically suffer with respect to bagging models). Finally, in the case of SHAP values, the intrinsic strategy of the method itself, based on the exploration of the full set of combinations among data patterns and related parameter space features, was insensitive to their potential variability.

\subsection{Evolutionary parameters}
The SEDs spanning the NIR to sub--mm range are widely recognized as valuable tools for assessing the evolutionary stage of a given object \citep{lada1984nature, allen2004infrared}. An important step for clump classification is the synergic exploitation of NIR/MIR and FIR/sub-mm data. This multi-wavelength approach is critical as it probes distinct physical components and processes inherent to star formation. 
While NIR/MIR observations allow us to penetrate some of the obscuring dust to identify embedded YSOs and trace warmer dust emission associated with protostars, FIR/sub-mm wavelengths are essential for characterizing the bulk properties of the cold, dense molecular gas and dust reservoirs from which stars form. In this perspective, the combination of these spectral windows provides a more complete understanding of clump evolutionary stages, mitigating biases inherent in single-wavelength studies and enabling a more robust assessment of their star-forming embedded stellar populations. 

Typically, the clump SED peak shifts progressively from wavelengths longer than 100~$\mu$m in the earlier-stage objects (where the infrared excess comes from radiation from the disk and envelope for the youngest objects, and average values of dust temperatures do not go beyond $\sim 15$ K, see for example \citealt{elia2017hi}) to shorter wavelengths as the object evolves. Among our selected features, the Herschel fluxes at 160~$\mu$m and 500~$\mu$m stand out. As reported in \citet{elia2017hi}, in the absence of a complete grey-body fit, the combination of these two fluxes are an excellent diagnostic for estimating the mean dust temperatures of clumps. 

As regards the colors, among the selected features we find the Herschel [250-350], and the [w4-70]. The latter has been widely used in literature in combination with other Herschel colors for diagnostic. For example, \cite{paladini2012spitzer} highlighted that it is useful for the segregation of the H\textsc{ii} regions with respect to the YSOs population with the color [160-250], and also  \cite{molinari2008evolution} emphasized the role of [24-70] to disentangle different kind of objects. 

As far as physical properties are concerned, in the last years different indicators has been useful for studying the evolutionary stages of clumps. Among the most known evolutionary indicators in the literature, there are the bolometric luminosity ($L_{\textrm{bol}}$) and its ratio to the mass (\textit{M}), namely the \lm. The \lm\ is a distance-independent quantity which is expected to rapidly increase during the accretion phase of star formation \citep{molinari2008evolution,elia2013first}. In our experiment, these two properties are not among the most important features. However, both $L_{\textrm{ratio}}$ and the surface density appear as important features. The $L_{\textrm{ratio}}$ is related to the bolometric luminosity, this being defined as the ratio between the bolometric luminosity and the luminosity at wavelengths longward of 350~$\mu$m. It is a distance-independent evolutionary indicator usually adopted, which is expected to be larger at more evolved star formation stages \citep{andre1993submillimeter}. For example, it has been investigated in studies about the low-mass star formation \citep{andre1999pre, maury2011formation}. In general, a significant emission observable at sub-mm wavelengths indicates an object in a very early stage of star formation \citep{elia2017hi}. Beside the $L_{\textrm{ratio}}$, the surface density (\textit{$\Sigma$}) is the parameter related to the mass of the clump. This is in fact obtained by dividing the mass of the clump by the area of a circle whose diameter is the linear diameter of the clump \citep{elia21}, thus it derives from both photometric measurements and the source linear sizes. It has been shown in \cite{elia2017hi} that more evolved sources are on average slightly smaller and denser than in the early stages of star formation, with a drop at increasing ages \citep{elia2017hi,giannetti2013physical}.

Another diagnostic is the bolometric temperature, $T_{\textrm{bol}}$, used in different works, such as \cite{strafella2015young}, for understanding the evolutionary stages of sources in Vela-D using \hg\ data. The $T_{\textrm{bol}}$ is often used with the $L_{\textrm{bol}}$ for building the corresponding $L_{\textrm{bol}}$ versus $T_{\textrm{bol}}$ diagram as a diagnostic tool for the characterization of YSOs from an evolutionary point of view \citep[for example,][]{strafella2015young}.

Lastly, the YSO density, $\delta_{YSO}$, is a feature selected in both the experiments. Coming from the \STEvMSCARS step, it is the percentage of YSOs within clumps, computed as the ratio of the number of YSOs to the total number of sources detected within the clump's spatial boundaries (encompassing all source types and including potential contaminants). Interestingly, our analysis reveals that  $\delta_{YSO}$ shows no significant correlation with commonly used physical or photometric parameters. This suggests that data-driven approaches can uncover observables that, while not grounded in physical modeling, encapsulate latent evolutionary information, offering a different perspective on clump evolution.

As previously discussed, the feature selection process primarily favored photometric features, likely because they trace dust emission and spectral signatures typical of embedded YSOs within clumps.
This outcome is likely influenced by the way the \textit{clump class} itself is defined. As explained in Sections~\ref{sec:clumppopulator}--\ref{sec:stemscars}, the class assigned to each clump is derived from the classification of the YSOs located within their elliptical area. The classification of individual YSOs is based on their fluxes and corresponding colors in the NIR/MIR range. Consequently, the \textit{clump class} is the average over the YSO classes in the clump and intrinsically tied to their NIR/MIR emission.
From this analysis, it appears that the NIR/MIR properties, by being inherently linked to the definition of the target variable, may overshadow the contribution of the physical properties in the feature selection process.
This does not necessarily imply that key physical properties, such as $L_{bol}$ and the \lm\ ratio, are irrelevant for tracing evolution. Rather, their exclusion may stem from other factors, such as strong correlations with selected photometric features, larger uncertainties in their estimates, or a reduced completeness across the sample. Additionally, their dynamic range or distribution may not be well captured by the selection algorithm, especially if it assumes linear relationships or normalized inputs. 
On the contrary, in the absence of features directly related to the definition of the \textit{clump class}, the algorithm relies entirely on structural and physical descriptors to discriminate among classes. While this shift in feature relevance is partly driven by the available input parameter space, it is particularly encouraging from a physical point of view: the selected features may describe the physical nature of the clumps, rather than simply reflecting how the clump class has been defined. In this sense, identifying this subset of relevant physical features could provide valuable hints to better understand the evolutionary processes at play in the formation and development of clumps.
Future work will aim to disentangle these effects and explicitly investigate how the selected photometric features are related to the underlying physical parameters, in order to build a more comprehensive picture of the evolutionary sequence.

\subsection{Statistical test analysis}
\begin{figure}[!hbpt]\centering

    \includegraphics[width=\columnwidth]{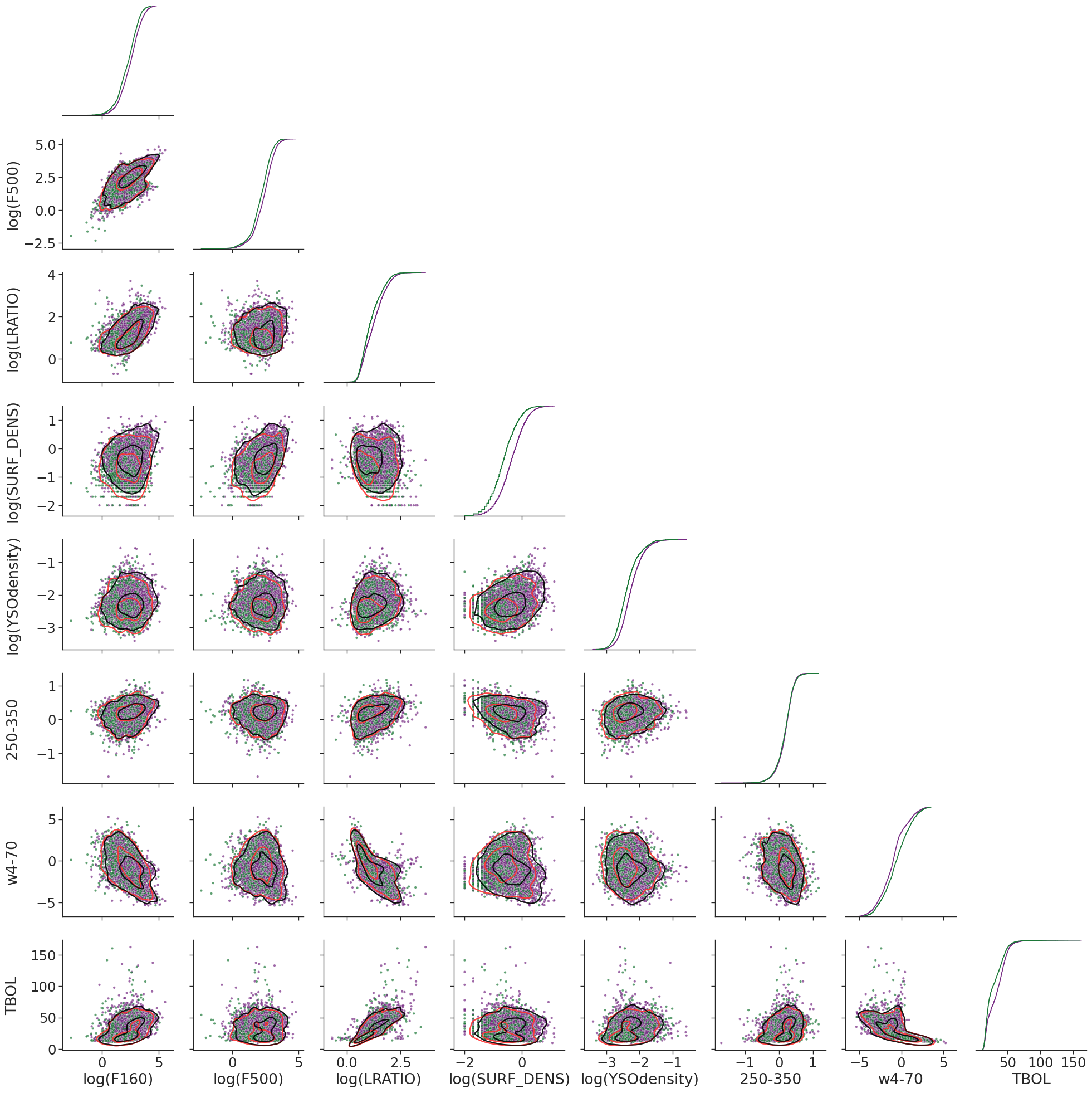}

    \definecolor{colorClass0}{HTML}{762a83}
    \definecolor{colorClass1}{HTML}{1b7837}

    \begin{tikzpicture}
        \node[draw, fill=colorClass0, minimum width=0.4cm, minimum height=0.4cm, opacity=0.3] at (0,0) {};
        \draw[black, thick] (0.3, 0) -- (0.7, 0); 
        \node at (1.3, 0) {Class 0};
        
        \node[draw, fill=colorClass1, minimum width=0.4cm, minimum height=0.4cm, opacity=0.3] at (2.8,0) {};
        \draw[red, thick] (3.1, 0) -- (3.5, 0); 
        \node at (4.1, 0) {Class 1};
    \end{tikzpicture}
    
    \caption{Grid of plots of the selected features for the experiment with the \textit{base PS}. In this grid, each plot corresponds to a pair of variables. Class 0 are plotted in violet, Class 1 in green. The contours represent the 20, 50 and 80 per cent levels of the density distribution of Class 0 (black) and Class 1 (red), respectively. Along the diagonal, the empirical cumulative distribution functions (ECDFs) of the given features are shown.}
    \label{fig:sel_feat_distribution}
    
\end{figure}
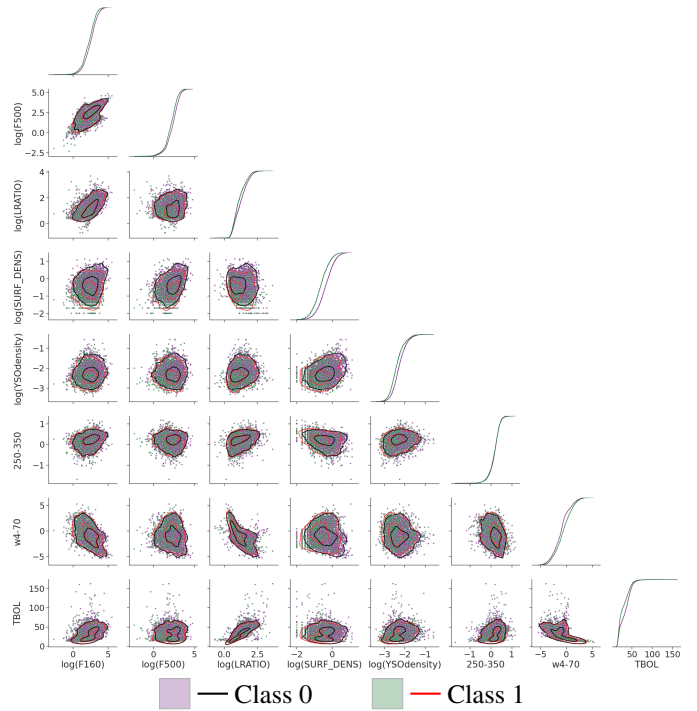

In Figure~\ref{fig:sel_feat_distribution}, a grid of plots of the the selected features is shown for providing a comprehensive overview of the data and enabling the identification of possible trends. 

First, it can be observed that while some pairs of features exhibit a relationship, albeit weak, others show no apparent correlation, with distributions suggesting independence between the variables. Second, the behavior of some features as a function of the \textit{clump class} shows significant overlap for certain features, whereas for others, the distributions appear to be quite distinguishable. For example, the distribution of [w4-70] is steeper for Class 0, assuming lower values than Class 1. In particular, we observe that high values of this color correspond to low values of $L_{\textrm{ratio}}$. This is expected, since in the early stages of star formation the sub-mm luminosity is higher, and it is difficult to have a strong emission in the MIR. However, the distinction between Class 0 and Class 1 is not as clear-cut as, for example, in the log(\textit{$\Sigma$}) vs [w4-70] plot. Generally, it can be observed that Class 0 exhibits higher values of \textit{$\Sigma$} compared to Class 1, as \textit{$\Sigma$} tends to be higher in the early stages of the star formation process and decreases as the process goes on \citep[e.g.][]{elia21}. Similarly, the $\delta_{YSO}$ exhibits the same behavior. It is important to note that these two parameters have been obtained independently. In fact, while the first is directly related to the mass of the clump \citep{elia2017hi}, the second is a statistical parameter derived from \STEvMSCARS and defined as the ratio between the number of YSOs in the clump and the total number of sources within the clump boundaries \citepalias{maruccia}. 

To determine whether the distributions of the selected features differ significantly between the two classes defined by the \textit{clump class}, we employed three statistical tests: the Kolmogorov-Smirnov (KS) test \citep{massey1951kolmogorov}, the Mann-Whitney U (MWU) test \citep{gibbon1992nonparametric}, and Linear Discriminant Analysis (LDA) \citep{mclachlan2005discriminant}. These tests provide complementary insights into the separability of the distributions. The KS test is a non-parametric test that evaluates whether two samples originate from the same continuous distribution by comparing their empirical cumulative distribution functions (ECDFs, represented also in Figure~\ref{fig:sel_feat_distribution}), making it particularly effective for detecting both location (along the \textit{x}-axis) and shape differences between distributions. The MWU test, on the other hand, is a non-parametric test that assesses whether one distribution tends to have systematically higher or lower values compared to the other, focusing primarily on differences in central tendency without assuming a specific distributional shape. Finally, LDA is not a statistical test but a supervised classification method, used to separate two or more classes in a space defined by multiple variables. Unlike the KS and MWU tests, LDA is specifically designed to address problems of separation and classification in multivariate contexts. In particular, it provides a quantitative measure of separability by constructing linear combinations of the features to maximize class distinction, offering a direct assessment of the discriminatory power of the variables. Together, these tests form a robust framework for evaluating the degree of separation between the two groups, each contributing unique strengths to the analysis: the KS test for overall distributional differences, the Mann-Whitney U test for central tendency shifts, and LDA for assessing classification performance. Results are summarized in Table~\ref{tab:stat_test_results}. 

\begin{table}[!hbpt]
\centering
\resizebox{\columnwidth}{!}{
\begin{tabular}{lclr}
\hline
\hline
\textbf{Feature}     & \textbf{KS Statistic} & \textbf{KS \textit{p}-value} & \textbf{MWU \textit{p}-value}  \\ \hline
$F_{\textrm{160}}$       & 0.1019                & $8.27 \times 10^{-14}$ & $< 10^{-8}$  \\
$F_{\textrm{500}}$       & 0.1217                & $1.64 \times 10^{-19}$ & $< 10^{-8}$  \\
$L_{\textrm{ratio}}$     & 0.1208                & $3.20 \times 10^{-19}$ & $< 10^{-8}$  \\
\textit{$\Sigma$}        & 0.2477                & $5.49 \times 10^{-80}$ & $< 10^{-8}$  \\
$\delta_{YSO}$           & 0.1724                & $8.77 \times 10^{-39}$ & $< 10^{-8}$  \\
$[250-350]$              & 0.0356                & $4.63 \times 10^{-2}$                 & $5.68 \times 10^{-2}$       \\
$[w4-70]$                & 0.1460                & $6.83 \times 10^{-28}$ & $< 10^{-8}$  \\
$T_{\textrm{bol}}$       & 0.1698                & $1.22 \times 10^{-37}$ & $< 10^{-8}$  \\ 
\hline
\end{tabular}
}
\caption{Statistical test results for feature separation adopting the \textit{clump class}. The first column reports the selected features for the experiment with the \textit{base PS}. The second and third columns show the distance between the two distributions and the probability of obtaining that distance by chance according to the KS test, while the fourth column those related to the MWU test.}
\label{tab:stat_test_results}
\end{table}

The statistical analysis performed on the selected variables highlights significant differences in the distributions when comparing the two classes identified by \textit{clump class}. The KS test reveals notable separability for nearly all variables, and this is evidenced by very low \textit{p}-values, which indicate that the null hypothesis of identical distributions can be confidently rejected. Indeed among these variables, \textit{$\Sigma$} stands out with the highest KS statistic of 0.2477 and a \textit{p}-value on the order of 10$^{-80}$, suggesting a particularly strong separation between the two classes. Similarly, $\delta_{YSO}$, $T_{\textrm{bol}}$, and [w4-70] demonstrate substantial separability, with KS statistics ranging from 0.1460 to 0.1724 and corresponding \textit{p}-values that remain far below any conventional significance threshold. The variable [250-350] presents a weaker signal, as indicated by a lower KS statistic of 0.0356 and a \textit{p}-value of 0.046, which, while still statistically significant, suggests that this feature contributes less to class differentiation compared to the others.

The MWU test further corroborates these findings, as it consistently returns \textit{p}-values close to zero for the majority of the variables, reinforcing the evidence that the distributions across the \textit{clump class} are significantly different. Only the variable [250-350] approaches the threshold of significance with a \textit{p}-value of approximately 0.057, aligning with the weaker signal observed in the KS test. These results collectively suggest that \textit{$\Sigma$}, $\delta_{YSO}$, and $T_{\textrm{bol}}$ are among the most discriminative features in separating the two classes, followed closely by $F_{\textrm{160}}$, $F_{\textrm{500}}$, and [w4-70].

The effectiveness of these features in capturing class separability is further supported by the LDA, which achieves an accuracy score of 0.6532. While this result does not indicate perfect separation, it demonstrates that the selected features contain meaningful discriminatory power when distinguishing between the two classes. The fact that LDA achieves this level of accuracy using linear combinations of the features suggests that the observed separability is not an artifact of the tests but reflects genuine differences between the two groups.

\section{Conclusions}\label{sec:conclusions}
In this work we studied the evolutionary path of star forming clumps in \hg\ through the adoption of a multi-step approach. First, we conducted feature engineering on data derived from several catalogs, from NIR to sub-mm, in order to build a broad parameter space able to trace the entire process of star formation, from the very early stage of the birth of YSOs to the classical pre-Main Sequence phase. Once the dataset was complete and processed to address any missing data and ensure normalization, we identified two distinct parameter spaces: the \textit{base PS}, made up of 16 features, and the \textit{enriched PS}, composed by 135 features. Following this, we applied a feature selection step, leveraging a combination of different algorithms (RFE + standard tree-based importance, RFE + SHAP importance, Boruta + standard tree-based importance, Boruta + SHAP importance) for improving the performance of our workflow, using the \textit{clump class} provided by \citetalias{maruccia} as target variable. For each experiment, we evaluated the results and determined the optimal combination for our parameter spaces. This process allowed us to define a reduced parameter space for each, consisting solely of the selected features: for the \textit{base PS}, 8 features have been selected as relevant, 43 for the \textit{enriched PS}. By adopting feature selection methods, we refined our models to focus solely on the most meaningful and relevant features. This strategy was essential for faithfully capturing the processes underlying star formation, ensuring a robust and insightful representation of this complex phenomenon. Finally, these four parameter spaces (the enriched, the base and the corresponding with only the selected features) have been used in the classification step of our workflow. This step validated the outcomes of the feature selection process. Main results are summarized below:

\begin{itemize}
    \item[(i)] The FS experiment with the \textit{enriched PS} revealed that the most significant features predominantly belong to the photometric group. It is worth noting that our method selects the most relevant features, which may result in similar features being included in the selection.

    \item[(ii)] The FS experiment with the \textit{base PS} suffers the scarcity of information related to the photometric features. This result was quite expected, due to the fact that the \textit{clump class} is related to the presence of YSOs within them, which in turn have been classified using their magnitudes at different wavelengths in the NIR/MIR. 

    \item[(iii)] Among the selected features, there are well-known indicators of the evolutionary stages of clumps in literature: the Herschel fluxes $F_{\textrm{160}}$ and $F_{\textrm{500}}$, the colors [250-350] and [w4-70], the ratio $L_{\textrm{ratio}}$ of the bolometric luminosity to the luminosity at wavelengths longward of 350~$\mu$m, bolometric temperature $T_{\textrm{bol}}$, the surface density \textit{$\Sigma$}, and the $\delta_{YSO}$.
    
    \item[(iv)] The classification experiment reveals interesting insights into the role of the selected features. While the \textit{base PS} lacked the necessary information to clearly distinguish between the classes (leading to notable contamination, particularly in Class 0), the feature selection process proved remarkably effective. By isolating the 8 most statistically significant variables, the model was able to retain its predictive performance even after dimensionality reduction. This outcome highlights that while the \textit{base PS} alone may lack the full range of information needed to clearly distinguish between the classes, the FS step effectively preserves key discriminative features (Table~\ref{tab:exp_metrics}, panels a and b). 

    \item[(v)] The combination of KS tests, MWU tests and LDA accuracy highlights a robust pattern of separation across the \textit{clump class} groups. Variables such as \textit{$\Sigma$}, $\delta_{YSO}$, and $T_{\textrm{bol}}$ emerge as the strongest candidates for characterizing the underlying differences between the two classes, while [250-350] shows more limited discriminatory capacity.

\end{itemize}

In summary, what emerges from our work is that some features lack a clear and consistent behavior in distinguishing the different evolutionary phases of the clumps. In particular, the NIR/MIR features do not appear to be strongly reflected in the evolutionary indicators of the clumps, which are primarily based on FIR features derived from their global emission. This result likely indicates that the global FIR/sub-mm emission of clumps and the radiation from the NIR/MIR resolved YSOs within them are not necessarily coeval but rather correspond to different stages of star formation \citepalias{maruccia}. These findings highlight the complex nature of star formation processes, where the physical conditions probed by different observational indicators can evolve asynchronously within the same clump environment.

This complex behavior obviously makes the problem of object classification complex and challenging. Traditional approaches to the classification of YSOs, in particular, have typically relied only on limited information, often related to color-color and color-magnitude diagrams. These include combination of \textit{Spitzer} and MIPS data complemented by \textit{J}, \textit{H} and \textit{K} from \twom\ \citep{gut08,gut09}, as well as \wise\ -based diagnostics combined with \twom\ \citep{koe12,koe14} or \textit{AKARI} \citep{pollo2010}, and so on. Although such approaches introduced a more formal statistical treatment, they remained constrained to low-dimensional parameter spaces and a limited set of features. 
A step toward more automated classification was introduced by \citet{toth2014akari}, who applied Quadratic Discriminant Analysis \citep[QDA, ][]{hastie2009elements} to a combination of \wise\ and \textit{AKARI} photometry, successfully recovering a high fraction of known YSO candidates with limited contamination. Similarly, \citet{marton2016all} applied Support Vector Machine \citep[SVM, ][]{cristianini2000introduction} trained on a dataset obtained combining Reliable \twom\ and \wise\ photometric data with \textit{Planck} dust opacity values, successfully identifying more known YSOs of the training sample than other methods based on color–color and magnitude–color selection.
In contrast, our study leverages a richer and more diverse dataset, combining all  available photometric information for identifying YSOs in the Galactic plane region 10\arcdeg$\leq \ell \leq$60\arcdeg, $|b| \leq 1^{\circ}$, including the \textit{Herschel} bands and the physical properties associated to the \hg\ clumps containing the YSOs. 
Our machine learning approach enables the integration of multiple photometric and physical features, capturing complex, possibly nonlinear relationships between observables and evolutionary classes. This facilitates a more objective, reproducible, and scalable classification process, which not only supports current models of clump evolution but also reveals the relative importance of diagnostic features that may be overlooked by traditional methods.

Although our analysis is based on \hg\ clumps, which are primarily associated with intermediate-to-high-mass star forming regions in the Galactic plane, the proposed methodology is general and can be applied to other star forming environments. The feature selection and classification framework does not rely on specific assumptions tied to \hg\ data and can, in principle, be adapted to different surveys. The inclusion of additional datasets in an experiment of this kind could, on the one hand, enrich the physical interpretation of the clump classification scheme, particularly when the new tracers differ significantly from the infrared continuum examined so far (for instance, data from line surveys or radio continuum observations). On the other hand, such an expansion towards other surveys would effectively constitute a new experiment, which might reveal that the newly introduced parameters are entirely independent of those identified in the previous analysis, or alternatively, that they add only limited relevance or novelty.

The advantage of performing a FS, besides its intrinsic capability to infer knowledge on the physics of the problem, is evident by comparing the statistical performance of the classification between the full and reduced parameter spaces. In fact, the comparable results in terms of purity and completeness demonstrates the presence of redundant information in the full case, useless from the classification point of view and highly impacting the experiment computation complexity. 

Finally, the interpretability of the resulting selected PS is intrinsically ensured by the use of SHAP method, developed in the framework of the Explainable Artificial Intelligence (XAI) methodology and proven to enhance the interpretability of its results in the context of data-driven approaches.
In particular, SHAP provides a unified and theoretically grounded approach to interpret model predictions by computing the marginal contribution of each feature based on cooperative game theory. This allows for a detailed attribution of the model’s output to individual input parameters, ensuring transparency and enabling a rigorous assessment of the physical relevance of the selected features. In this context, SHAP serves not only as a post-hoc explanation tool but also as a guide for domain-driven interpretation, strengthening the methodological robustness and scientific interpretability of the proposed approach.

\section*{Acknowledgements}
The authors thank the anonymous referee for the valuable comments and suggestions that have improved the quality of this manuscript. The authors acknowledge funding by the European Research Council via the ERC Synergy Grant ``ECOGAL – Understanding our Galactic ecosystem: From the disk of the Milky Way to the formation sites of stars and planets'' (project ID 855130). 
YM, SC, GR acknowledge the Spoke 3 ``Astrophysics and Cosmos Observations'' of the ICSC–Centro Nazionale di Ricerca in High Performance Computing, Big Data and Quantum Computing and hosting entity, funded by European Union–Next GenerationEU. YM expresses gratitude to Professor Francesco Strafella for his helpful comments and feedback, which greatly strengthened the overall manuscript.
SC and GR acknowledge support from PRIN MUR 2022 (20224MNC5A), ``Life, death and after-death of massive stars'', funded by European Union – Next Generation EU.  
MB, SC and GR acknoweldge the ASI-INAF TI agreement, 2018-23-HH.0 ``Attività scientifica per la missione Euclid - fase D''.




\bibliographystyle{elsarticle-harv} 
\bibliography{biblio.bib}






\end{document}